\documentclass[12pt]{article}
\usepackage{a4wide}
\title{{\bf A neural-network-like quantum \\ information processing system}}
\author{Mitja Peru\v s and Horst Bischof \thanks{Institute for Computer
Vision and Graphics, Graz University of Technology; Inffeldgasse
16, 2.OG; A-8010 Graz, Austria, EU \ / \ perus@icg.tu-graz.ac.at \
/ \ www.icg.tu-graz.ac.at/$\sim$perus \ / \ Fax +43-316-873-5050 \
\ Tel -5029}}
\date{} \textheight=24.3cm \textwidth=16.5cm \voffset=-2cm
\begin{document}
\maketitle
\begin{abstract}
The Hopfield neural networks and the holographic neural networks
are models which were successfully simulated on conventional
computers. Starting with these models, an analogous fundamental
quantum information processing system is developed in this
article. Neuro-quantum interaction can regulate the
"collapse"-readout of quantum computation results. This paper
is a comprehensive introduction into associative processing
and memory-storage in quantum-physical framework. 
\end{abstract}
{\bf 1. INTRODUCTION} \\
\\
The aim of this paper is to provide a general introduction and a
short coherent overview of recent research of quantum
Hopfield-like information processing --- to show quantum
information theorists where they have common issues with
neural-net modelers.

 In Peru\v s (1996, 1997, 1998) it was systematically
presented how and where the mathematical formalism of associative
neural network models (Hopfield, 1982; Amit, 1989) and synergetic
computers (Haken, 1991) is {\it analogous} to the mathematical
formalism of quantum theory. In this paper some of these analogies
will be used in an original presentation of some information
processing capabilities of quantum systems in nature -- i.e., not
in artificial devices, although the latter option is also open. In
parallel to ours, two interesting models of artificial quantum
neural networks have been independently developed by Bonnell and
Papini (1997) and by Zak and Williams (1998).

This contribution attempts to describe one of the simplest, but
fundamental, quantum Hopfield-like information processing
"algorithms". We have chosen this "algorithm", simple but
nevertheless effective, technically realizable and biophysically
plausible enough (Pribram, 1991, 1993), as a convenient one for
presenting neural-network-like quantum information dynamics. By
saying Hopfield-like we mean a system that is based on the
Hopfield model of neural nets or spin glass systems, respectively
(Dotsenko, 1994; Geszti, 1990). The Hopfield "algorithm" has been
extensively tested by our computer simulations using various
concrete data sets (Peru\v s, 2002; Peru\v s and E\v cimovi\v c,
1998; Haykin, 1994). Among others, we have effectively realized
parallel-distributed content-addressable memory, selective
associative reconstruction or recognition of patterns memorized in
a compressed form, and even some limited capability for
predictions based on a learned data set. We have analyzed how the
results are dependent on the correlation structure of a specific
set of input patterns, and conditions for successful processing
(Peru\v s, 2002; see also Peru\v s and Dey, 2000).
\\
\\
\\
{\bf 2. ASSOCIATIVE NEURAL NET MODEL} \\

In the Hopfield associative network model, emergent collective
computation or learning is globally regulated by minimization of
the spin-glass-like Hamiltonian energy function (Amit, 1989;
Dotsenko, 1994)
$$ H \ = \ - \frac{1}{2} \sum_{i=1}^N \sum_{j=1}^N \ J_{ij} \ q_i
\ q_j \ = \ - \frac{1}{2} \sum_{i=1}^N \sum_{j=1}^N \sum_{k=1}^P \
v_i^k \ v_j^k \ q_i \ q_j \eqno(1) .$$ $q_i$ is the actual
activity of the $i^{th}$ neuron. There are $N$ neurons in the
network: $ \vec{q} = (q_1,...,q_N) $. $v_i^k$ is the activity of
the neuron $i$ when taking part in encoding the $k^{th}$ memory
pattern ($k$ is pattern's {\it superscript} index). The process of
gradient-descent of energy function (1) is a result of
interactions between the system of neurons, described by
$\vec{q}$, and the system of "synaptic" connections, described by
weights $J_{ij}$ which are elements of the memory matrix {\bf J}
(Peru\v s and E\v cimovi\v c, 1998).

In an equivalent, local "interactional" description, the dynamical
equation for neuronal activities $$ q_i (t_2 = t_1 + \delta t) \ =
\ \sum_{i=1}^N \ J_{ij} \ q_j (t_1) \ \ \ \ \ \ \ or \ \ \ \ \ \ \
\vec{q} (t_2) = {\bf J} \ \vec{q} (t_1) \eqno(2) $$ is coupled
with the Hebb dynamical equation for "synaptic" connections $$
J_{ij} \ = \ \sum_{k=1}^P \ v_i^k \ v_j^k \ \ \ \ \ \ \ or \ \ \ \
\ \ \ {\bf J} = \sum_{k=1}^P \ \vec{v}^k \ (\vec{v}^k)^T \eqno(3).
$$ This system of equations is enough for realizing
parallel-distributed information processing of input data. It is
the core of one of the simplest "algorithms" useful for
theoretical brain modeling (Amit, 1989) as well as for automatic
empirical modeling of concrete experimental data sets (Haykin,
1994).

Input-data vectors $\vec{v}^k$ can be inserted into the system of
neurons $\vec{q}$ iteratively, or can be put in the very beginning
simultaneously into the Hebb matrix {\bf J} which contains all
"synaptic" weights $J_{ij}$ (equation (3)). Let us rewrite the
system (2) and (3) into continuous description of activities of
neurons and synapses at position $\vec{r}$ and time $t$: $$ q
(\vec{r}_2, t_2 = t_1 + \delta t) = \int J (\vec{r}_1, \vec{r}_2)
\ q (\vec{r}_1, t_1) \ d\vec{r}_1 \eqno(2b) $$ $$ J (\vec{r}_1,
\vec{r}_2) \ = \ \sum_{k=1}^{P} \ v^k (\vec{r}_1) \ v^k
(\vec{r}_2) \eqno(3b) . $$

The memory recall is done by $ \vec{q}_{output} = {\bf J}
\vec{q}_{input} = {\bf J} \vec{q} \, ' $.
Variable with a prime means that its quantitative value is close to
the variable without prime, i.e. $ q' \doteq q $.
This can be analyzed by
$$ q (\vec{r}_2, t_2) = \int J (\vec{r}_1, \vec{r}_2)
\ q' (\vec{r}_1, t_1) \ d\vec{r}_1 = \int \left(
\sum_{k=1}^P v^k (\vec{r}_1) \ v^k (\vec{r}_2) \right)
\ q' (\vec{r}_1, t_1) \ d\vec{r}_1 = $$
$$ = \left( \int v^1 (\vec{r}_1)
\ q' (\vec{r}_1, t_1) \ d\vec{r}_1 \right) v^1 (\vec{r}_2) +
 \left( \int v^2 (\vec{r}_1)
\ q' (\vec{r}_1, t_1) \ d\vec{r}_1 \right) v^2 (\vec{r}_2) + ...
$$ $$ + \left( \int v^P (\vec{r}_1) \ q' (\vec{r}_1, t_1) \
d\vec{r}_1 \right) v^P (\vec{r}_2) = $$ $$ = \ A \ v^1 (\vec{r}_2)
\ + \ B \ \ \ \ \ \ \ where \ \ A \ \doteq \ 1 \ \ ('signal'), \ B
\ \doteq \ 0 \ \ ('noise') \eqno(4) $$ or in another description
$$ q (\vec{r}, t) = \sum_{k=1}^P C'^k (t) v^k (\vec{r}) =
\sum_{k=1}^P \left( \int v^k (\vec{r}) q' (\vec{r}, t) d\vec{r}
\right) v^k (\vec{r}) = $$ $$ = \left( \int v^1 (\vec{r}) q'
(\vec{r}, t) d\vec{r} \right) v^1 (\vec{r}) + \left( \int v^2
(\vec{r}) q' (\vec{r}, t) d\vec{r} \right) v^2 (\vec{r}) + ... $$
$$ + \left( \int v^P (\vec{r}) q' (\vec{r}, t) d\vec{r} \right)
v^P (\vec{r}) = $$ $$ = \ A \ v^1 (\vec{r}) \ + \ B \ \ \ \ \ \ \
where \ \ A \ \doteq \ 1 \ \ ('signal'), \ B \ \doteq \ 0 \ \
('noise') \eqno(5). $$ The first row of equalities in eq. (5)
follows, among others, from the neurosynergetic model by Haken
(1991). In eq. (4) and eq. (5) we had to choose such an input
$\vec{q} \, '$ that is {\it more similar} to $\vec{v}^1$, for
example, than to any other $\vec{v}^k, k \ne 1$. At the same time,
the input $\vec{q} \, '$ should be {\it almost orthogonal} to all
the other $\vec{v}^k, k \ne 1$. In this case, $\vec{q}$ converges
to the memory pattern-qua-attractor $\vec{v}^1$, as it is shown in
the last row of eq. (4) and in the last row of eq. (5). Thus,
$\vec{v}^1$ is recalled. \\
\\
\\
{\bf 3. HOLOGRAPHIC NEURAL NET MODEL} \\

If we introduce the {\it oscillatory} time-evolution of neuronal
activities, we get more biologically plausible dynamics (Baird,
1990; Schempp, 1994; Haken, 1991; Kapelko and Linkevich, 1996).
Neural net variables then become complex-valued. This is more akin
to quantum dynamics. However, in quantum theory, complex-valued
formalism is essential, but in the network of coupled oscillatory
neurons complex-valued formalism is just mathematically more
convenient.

Simulated holographic neural system by Sutherland (1990) realizes
efficient Hopfield-like information processing which incorporates
oscillatory activities. A Hebb-like "interference" of sequences of
input--output pairs (indexed by $k$), each combining an input
vector $ \vec{s}^k = (s_1^k e^{i \theta_1^k},..., s_N^k e^{i
\theta_N^k}) $, $i = \sqrt{-1}$, and an output vector $ \vec{o}^k
= (o_1^k e^{i \varphi_1^k},..., o_M^k e^{i \varphi_M^k}) $,
analogous to eq. (3), is made: $$ {\bf J} = \sum_{k=1}^P \vec{o}^k
(\vec{s}^k)^{\dagger} \Longleftrightarrow J_{hj} = \sum_{k=1}^P
s_h^k o_j^k e^{i (\varphi_j^k - \theta_h^k)} \eqno(6). $$ Every
$J_{hj}$ encodes a sequence of local input--output amplitude
correlations ($ s_h o_j $) and corresponding phase differences ($
\varphi_j^k - \theta_h^k $) in local input and output oscillatory
dynamics. Thus, the phase differences appear in the elements of
the memory matrix {\bf J} which represents the "hologram". If
$\vec{s}^k$ and $\vec{o}^k$ are defined as two parts of a
concatenated vector $\vec{v}^k$ (occupying its "upper" and "lower"
set of components), then we have the Hebbian "self-interference",
like in eq. (3). Usually, during learning, inputs and output
(indexed by $k=1,...,P$) are "interfered" in a time sequence --
each pair $\vec{s}^k$ and $\vec{o}^k$ corresponds to a discrete
time $t_k$.

A pattern can be reconstructed (analogously to equations (4) or
(5)) from the "neural hologram" using a recall-key $\vec{s} \, '$:
$$ \vec{o} \, ' = {\bf J} \vec{s} \, ' \ \Longleftrightarrow \
\sum_{h=1}^N J_{hj} s'_h e^{i \theta'_h} = \sum_{h=1}^N
\sum_{k=1}^P s_h^k o_j^k e^{i (\varphi_j^k - \theta_h^k)} s'_h
e^{i \theta'_h} \doteq o_j^1 e^{i \varphi_j^1} \eqno(7). $$ This
is valid if we choose a recall-key that is {\it similar} to one of
the learned inputs, say $\vec{s}^1$. Thus, $s'_h \doteq s_h^1$ and
$\theta'_h \doteq \theta_h^1$, for all $h$. In such a case, $ s'_h
s_h^1 \doteq 1 $ (if they are normalized) and $ e^{i \theta'_h}
e^{-i \theta_h^1} \doteq 1 $. Other terms (with $k \ne 1$) are
relatively very small ('noise'). (See Sutherland (1990) for
comprehensive presentation.)

Holographic associative memories are implemented optically,
acoustically, quantum-elect\-ro\-ni\-cal\-ly and
quantum-biologically (Psaltis {\it et al.}, 1990; Psaltis and Mok,
1995; Nobili, 1985; Schempp, 1994). This supports our view that
Hopfield-like associative dynamics with Hebbian learning can be
implemented in various (bio)physical (see Jibu {\it et al.}, 1996)
and especially quantum systems (Bonnell and Papini, 1997; Zak and
Williams, 1998; Nishimori and Nonomura, 1996; Ma {\it et al.},
1993), including coherent states, quantum dots, quantum
information channels (Alicki, 1989; Ohya, 1989). Quantum
holography (e.g., Abouraddy
{\it et al.}, 2001) offers the most natural implementation. \\
\\
\\
{\bf 4. QUANTUM ASSOCIATIVE NET MODEL} \\

So far, we have discussed the Hopfield-like neural net or spin
glass model. Now we will present its quantum relative. From the
very beginning, let us put attention to the following
correspondence scheme between the neural (left) and quantum
variables (right) ("$\Leftrightarrow$" means "corresponds to"): $$
q \Leftrightarrow \Psi, \ \ \ q' \Leftrightarrow \Psi' , \ \ \ v^k
\Leftrightarrow \psi^k, \ \ \ J \Leftrightarrow G \ \ or \ \ {\bf
J} \Leftrightarrow {\bf G} , \ \ \ C^k \Leftrightarrow c^k $$ $$
equation \ (2b) \Leftrightarrow equation \ (8), \ \ \ (3b)
\Leftrightarrow (9), \ \ \ (4) \Leftrightarrow (10), \ \ \ (5)
\Leftrightarrow (11) $$ The phase $\varphi$ is hidden in the
exponent of the wave-function $\Psi$ which describes the state of
the quantum system: $ \Psi (\vec{r}, t) = A (\vec{r}, t) exp (
\frac{i}{\hbar} \varphi (\vec{r}, t) ) $; $A$ is the amplitude.

The equations in pairs are {\it mathematically equivalent},
because the {\it collective} dynamics in neural and quantum
complex systems are similar, in spite of different nature of
neurons and their connections on one hand, and quantum "points"
$\Psi (\vec{r})$ and their "interactions" described by $ G
(\vec{r}_1, \vec{r}_2 ) $ on the other.

The quantum Hopfield-like network model combines the dynamical
equation for the quantum state (Feynman and Hibbs, 1965; Derbes,
1996) $$ \Psi (\vec{r}_2, t_2) = \int \int \ G (\vec{r}_1, t_1,
\vec{r}_2, t_2) \ \Psi (\vec{r}_1, t_1) \ d\vec{r}_1 \ dt_1 \ \ \
\ \ or \ \ \ \ \ \Psi (t_2) = {\bf G} \ \Psi (t_1) \eqno(8) $$ and
the expression for the parallel-distributed interactive
transformation of the quantum system (Feynman and Hibbs, 1965) $$
G (\vec{r}_1, t_1, \vec{r}_2, t_2) = \sum_{k=1}^{P} \psi^k
(\vec{r}_1, t_1)^{\ast} \ \psi^k (\vec{r}_2, t_2) \ \ \ \ \  or \
\ \ \ \ G (\vec{r}_1, \vec{r}_2) = \sum_{k=1}^{P} \psi^k
(\vec{r}_1)^{\ast} \ \psi^k (\vec{r}_2) \eqno(9). $$ Note that
expression (9), i.e. for $ G (\psi^k (\vec{r}_1, t_1), \psi^k
(\vec{r}_2, t_2)) $, presents the kernel of eq. (8) (cf., Vapnik,
1998). The system (8) and (9) is just the usual Schr\"{o}dinger
propagation reinterpreted for associative processing and
measurement-like
readout. \\

We want to {\it encode some information} in eigenfunctions
$\psi^k$. Then $\psi^k$ would become quantum codes of patterns,
although not necessarily geometrically isomorphic to some external
patterns. It may not be possible to encode information in
wave-functions $\Psi$, or $\psi^k$, in the same sense (by the same
directly decodable way, respectively) that information is encoded
in neural-net state-vectors $\vec{q}$, or $\vec{v}^k$, for two
reasons. First, any natural (non-model) network may initially be
in some "natural" state, i.e. eigenstate, of its own. Second,
$\vec{q}$, or $\vec{v}^k$, are in principle directly observable,
but $\Psi$, or $\psi^k$, are (in general) not -- not even in
principle, as long as they remain quantum. Therefore, we act as
follows.

By a classical interaction or perturbation on an appropriate
quantum system, we force the quantum network into a state $\Psi$
which "implicitly reflects" our external influences (inputs), i.e.
it is input-modulated. As soon as such a state $\Psi$ stabilizes,
becoming an eigenstate, $\psi^k$ ($k=1$), we can continue to
"insert" (simultaneously or sequentially) other
information-encoding states ($k=2,...,p$). All these eigenstates
$\psi^k$ interfere as prescribed by equation (9) and get thus {\it
stored} in {\bf G}. {\it Quantum holography} (e.g., Abouraddy {\it
et al.}, 2001) is an example which demonstrates how this could be
realized, with plane waves $ \Psi = A \ e^{\frac{i}{\hbar}
\varphi} $ or wavelets (Schempp, 1994), without extensive
artificial effords. Moreover, fast-developing specially-designed
encoding / decoding (measurement) devices (e.g., Weinacht {\it et
al.}, 1999) enable enormous additional possibilities.

Now we want to {\it retrieve a pattern from memory}. In our
quantum-net model we are not interested in $\Psi$, like we are not
interested in $\vec{q}$ in our neural-net model. Our final result
will directly be a single "post-measurement" information-encoding
eigenstate $\psi^k$ (say $k=1$), or $\vec{v}^k$, respectively. We
cannot observe $\Psi$ (except in net-simulations by stopping the
program) and we {\it need not} observe $\Psi$ (or $\vec{q}$), but
we wait until the "measurement" (i.e., pattern-recall which is
equivalent to the wave-function "collapse") is triggered by our
final new input $\Psi'$ or $\vec{q}\, '$. Then, the standard
quantum {\it observables} $O$ (corresponding to: ${\bf \hat{O}}
\psi = \lambda_O \psi$), e.g. spin states, can reveal the
reconstructed pattern-encoding eigenstate $\psi^k$ \ ($k=1$).
Namely, since the output is similar to the input (that we know!),
the eigenvalue ($\lambda_O$) information can be sufficient for
knowing the output. Moreover, if the final $\Psi = \psi^k \ (k=1)$
is / becomes {\it classical} (like the inputs may well be), then
obtaining complete knowledge about the output pattern, encoded in
$\psi^k \ (k=1)$, is at least in some cases (e.g., optical) {\it
relatively straight-forward} (e.g., like {\it seeing} the image
reconstructed from a hologram). Beside quantum holography, an
alternative fast-developing technique is quantum tomography for
reconstruction of eigenstates $\psi^k$ (D'Ariano {\it et al.},
2000). So, our information-processing result {\it can be
extracted} from $\psi^k$ using new quantum-optical (and
computer-aided) techniques for measurement of observables or
for quantum-holographic-(like) wavefront reconstruction. \\

Let's {\it analyze the memory and how a pattern is retrieved from
it}. Quantum holography is our primary suggestion. If
eigenfunctions $\psi^k$ implicitly encode patterns presented to
the net, then matrix {\bf G} describes the {\it quantum memory}.
The propagator expression $G$ in eq. (9), which acts as a
projector during the pattern-recall (measurement) process, is
related to the usually-used Green function $\tilde{G}$ (e.g.,
Bjorken and Drell, 1964/65) by $ G = -i \tilde{G} $.

If we, in eq. (9) which looks Hebbian, expose the phases $\varphi$
explicitly, using $ \Psi = A \, exp ( i \varphi ) $, we get an
expression which is the {\it {\bf quantum phase}-Hebb learning
rule}: $$ G ( \vec{r}_1, t_1, \vec{r}_2, t_2 ) = \sum_{k=1}^P A^k
( \vec{r}_1, t_1 )^{\ast} A^k ( \vec{r}_2, t_2 ) e^{ - i ( \varphi
( \vec{r}_2, t_2 ) - \varphi ( \vec{r}_1, t_1 ) ) } \eqno(9b) .
$$
This describes the memory encoding which is two-fold: it is both
in amplitude-correlations $ \sum_{k=1}^P A_k (\vec{r}_1, t_1) A_k
(\vec{r}_2, t_2) $ (Hebb rule) and in phase-differences $ \delta
\varphi_k = \varphi_k (\vec{r}_2, t_2) - \varphi_k (\vec{r}_1,
t_1) $.

The difference between the rule (9b) and a non-quantum phase-Hebb
rule is that in eq. (9b) phases $\varphi$ are quantum phases ---
i.e., Planck's constant $h$ is hidden in the exponent (but the
usual notation $ \hbar = \frac{h}{2 \pi} = 1 $ is used now).

The {\it quantum memory retrieval} ($ \Psi_{output} = {\bf G}
\Psi' $) is most-directly realized by the input-triggered,
non-unitary {\it wave-function "collapse"}:
$$ \Psi (\vec{r}_2, t_2 = t_1 + \delta t) = \int G (\vec{r}_1,
\vec{r}_2) \ \Psi' (\vec{r}_1, t_1) \ d\vec{r}_1 = \int \left(
\sum_{k=1}^{P} \psi^k (\vec{r}_1)^{\ast} \psi^k (\vec{r}_2)
\right) \ \Psi' (\vec{r}_1, t_1) \ d\vec{r}_1 = $$ $$ = \left(
\int \psi^1 (\vec{r}_1)^{\ast} \Psi' (\vec{r}_1, t_1) d\vec{r}_1
\right) \psi^1 (\vec{r}_2) + \left( \int \psi^2 (\vec{r}_1)^{\ast}
\Psi' (\vec{r}_1, t_1) d\vec{r}_1 \right) \psi^2 (\vec{r}_2) + ...
$$ $$ + \left( \int \psi^P (\vec{r}_1)^{\ast} \Psi' (\vec{r}_1,
t_1) d\vec{r}_1 \right) \psi^P (\vec{r}_2) = $$ $$ = \ A \ \psi^1
(\vec{r}_2) \ + \ B \ \ \ \ \ \ \ where \ \ A \ \doteq \ 1 \ \
('signal'), \ B \ \doteq \ 0 \ \ ('noise') \eqno(10) $$ or in
another description $$ \Psi (\vec{r}, t) = \sum_{k=1}^P c'^k (t)
\psi^k (\vec{r}) = \sum_{k=1}^P \left( \int \psi^k
(\vec{r})^{\ast} \Psi' (\vec{r}, t) d\vec{r} \right) \psi^k
(\vec{r}) = $$ $$ = \left( \int \psi^1 (\vec{r})^{\ast} \Psi'
(\vec{r}, t) d\vec{r} \right) \psi^1 (\vec{r}) + \left( \int
\psi^2 (\vec{r})^{\ast} \Psi' (\vec{r}, t) d\vec{r} \right) \psi^2
(\vec{r}) + ... $$ $$ + \left( \int \psi^P (\vec{r})^{\ast} \Psi'
(\vec{r}, t) d\vec{r} \right) \psi^P (\vec{r}) = $$ $$ = \ A \
\psi^1 (\vec{r}) \ + \ B \ \ \ \ \ \ \ where \ \ A \ \doteq \ 1 \
\ ('signal'), \ B \ \doteq \ 0 \ \ ('noise') \eqno(11). $$ In eq.
(10) and eq. (11) we had to choose such an "input" $\Psi'$ that is
{\it more similar} to $\psi^1$, for example, than to any other
$\psi^k, k \ne 1$. At the same time, the "input" $\Psi'$ should be
{\it almost orthogonal} to all the other $\psi^k, k \ne 1$. In
this case, $\Psi$ converges to the quantum "pattern-qua-attractor"
$\psi^1$, as it is shown in the last row of eq. (10) and in the
last row of eq. (11). Thus, the memory pattern $\psi^1$ is
recalled (measured). If the condition, well known from the
Hopfield model simulations, that "input" must be similar to one
stored pattern (at least more than to other stored patterns) is
not satisfied, then there is no single-pattern
recall.  \\
%
%
%
\\
\\
{\bf 5. DISCUSSION} \\
\\
The system of quantum equations (8) and (9) is {\it similar,
according to their mathematical structure and coupling}, to the
system of neural-net equations (2) and (3). Because we are certain
that the neural system (2) and (3) realizes efficient information
processing, we have taken the similar system of equations from the
quantum formalism in order to discover quantum Hopfield-like
associative information dynamics.

There is a difference between the neural "algorithm" (2)---(5) and
the quantum "algorithm" (8)---(11): Neural variables like
$\vec{q}, \vec{v}^k$ and {\bf J} in equations (2)---(5) are
real-valued, but quantum variables like $\Psi$, $\psi^k$ and {\bf
G} are complex-valued. Important implications of this fact are
discussed in detail in Peru\v s (1996, 1997).

Anyway, (quantum) holography shows (Abouraddy {\it et al.}, 2001;
Psaltis {\it et al.}, 1990, 1995) that the quantum neural-net-like
information processing outlined here {\it is} realizable.
Moreover, it seems that neural networks and quantum networks
cannot necessarily be treated as complex systems with similar, but
{\it independent}, collective dynamics. There are strong
indications (Pribram, 1991, 1993) that biological neural networks
essentially cooperate with quantum networks in the brain.
Mediators are probably synapto-dendritic and microtubular nets.
All these networks constitute a sort of fractal-like multi-level
information processing (Peru\v s, 1996, 1997). The wave-function
collapse (a quantum sort of pattern recognition -- remember eq.
(10)) can be triggered by the system's interaction with
environment (see Zurek, 1991; Brune {\it et al.}, 1996). It seems
that, in the brain, neural networks sensing the environment
trigger the wave-function collapse and thus transform the quantum
complex-valued, probabilistic dynamics into the neural (classical)
real-valued, deterministic dynamics (Peru\v s, 1996, 1997; Peru\v
s and Dey, 2000). A consequence of the wave-function collapse is
that the quantum network becomes more neural-net-like, e.g. the
observable "activities of a network of quantum points ('neurons')"
are real-valued.

It is true that holographic (Sutherland, 1991) and other
oscillatory neural networks (e.g., Baird, 1991), too, do not
realize the deep essence of quantum dynamics, i.e. the
EPR-manifested
non-local interconnectedness and indivisibility, called
entanglement. But these features are broken and dynamics is thus
discretized during quantum measurements. The implicit
parallel-distributed quantum dynamics is useful for computation,
but it has to be collapsed during the readout (or memory recall)
process in order to obtain results of computation. These essential
quantum features, manifested in the complex-valued formalism,
could be harnessed or partly eliminated in order to realize
neural-net-like information processing. Although it is useful to
harness quantum superpositional multiplicity for computational
purposes (Steane, 1998; Scarani, 1998) as much as possible, it is
practically unavoidable to collapse the quantum wholeness during
the readout (measurement) of results. For this, neuro--quantum
(classical--quantum) cooperation, as manifested in the
collapse-readout, seems useful for the brain as well as for
hypothetical neuro-quantum computers (Kak, 1995). The advantages
of quantum information dynamics, i.e. miniaturization, speed,
computational and memory capacity, are preserved in that case.

Detailed technical (or biological) realization of encoding of
information into eigenfunctions $\psi^k$ and the readout of
results of quantum pattern-reconstruction is still an open and
advancing field of research. For example, encoding and decoding
(readout) methods using laser or NMR devices, developed for
quantum computers (Ahn {\it et al.}, 2000; Weinacht {\it et al.},
1999; Berman {\it et al.}, 1998; Jones {\it et al.}, 2000; Cirac
and Zoller, 1995), could partially be used. There are numerous
complex systems which are candidates for implementation of
collective computation which "follows algorithms" similar to the
one described here (Peru\v s,
1996, 1997; Pribram, 1991, 1993). \\

A fundamental quantum information processing "algorithm"
(8)---(11) was presented. It was constructed following the
Hopfield and holographic neural network models which process
information efficiently as tested by numerous computer simulations
(e.g., Peru\v s and E\v cimovi\v c, 1998; Amit, 1989; Sutherland,
1991).
It was shown how can the Hopfield-like associative information
processing and content-addressable memory, in principle, be
realized in "natural" quantum systems, i.e. without need for
special devices, except for encoding and decoding. This has
important consequences for foundations of physics and informatics
as well as for cognitive neuroscience. One of many possible
applications, i.e. quantum pattern-recognition, is presented at:
quant-ph/0303092. Concrete possibilities of quantum implementation
are also discussed there. \\
\\
\\
\small
{\bf ACKNOWLEDGEMENTS} \\
\\
EU Marie-Curie fellowship (HPMF-CT-2002-01808), GAMM and EIU
grants, and discussions with Professors L.I. Gould, K.H. Pribram,
W. Sienko, A.O. \v Zupan\v ci\v c, A.A. Ezhov and
J. Glazebrook are gratefully acknowledged by M.P. \\
\\
\\
{\bf REFERENCES} \\
\\
\small Abouraddy, A., {\it et al.} (2001). Quantum holography.
{\it Optics Express}, {\bf 9}, 498.
\\
Ahn, J., Weinacht, T.C., and Bucksbaum, P.H. (2000). {\it
Science}, {\bf 287}, 463.
\\
Alicki, R. (1997). {\it Open Sys. \& Information Dyn.}, {\bf 4},
53.
\\
Amit, D. (1989). {\it Modeling Brain Functions (The World of
Attractor Neural Nets)}, Cambridge Univ. Press, Cambridge.
\\
Baird, B. (1990). {\it Physica D}, {\bf 42}, 365.
\\
Berman, G.P., Doolen, G.D., Mainieri, R., and Tsifrinovich, V.I.
(1998). {\it Introduction to Quantum Computers}, World Scientific,
Singapore.
\\
Bjorken, J.D., and Drell, S.D. (1964/65). I: {\it Relativistic
Quantum Mechanics} / II: {\it Relativistic Quantum Fields},
McGraw-Hill, New York.
\\
Bonnell, G., and Papini, G. (1997). Quantum Neural Network. {\it
Int. J. Theor. Phys.}, {\bf 36}, 2855.
\\
Brune, M., {\it et al.} (1996). {\it Phys. Rev. Lett.}, {\bf 77},
4887.
\\
Cirac, J.I., and Zoller, P. (1995). {\it Phys. Rev. Lett.}, {\bf
74}, 4091.
\\
Derbes, D. (1996). {\it Amer. J. Phys.}, {\bf 64}, 881.
\\
D'Ariano, G., L. Maccone, and M. Paris (2000). {\it Phys. Lett.
A}, {\bf 276}, 25-30.
\\
Dotsenko, V. (1994). {\it The Theory of Spin Glasses and Neural
Networks}, World Scientific, Singapore.
\\
Feynman, R.P., and Hibbs, A.R. (1965). {\it Quantum Mechanics and
Path Integrals}, McGraw-Hill, New York.
\\
Geszti, T. (1990). {\it Physical Models of Neural Networks}, World
Scientific, Singapore.
\\
Haken, H. (1991). {\it Synergetic Computers and Cognition (A
Top-Down Approach to Neural Nets)}, Springer, Berlin.
\\
Haykin, S. (1994). {\it Neural Networks}, MacMillan, New York.
\\
Hopfield, J.J. (1982). {\it Proceed. Natl. Acad. Sci. USA}, {\bf
79}, 2554.
\\
Jibu, M., Pribram, K.H., and Yasue, K. (1996). {\it Int. J. Modern
Phys.}, {\bf 10}, 1735.
\\
Kak, S.C. (1995). {\it Information Sciences}, {\bf 83}, 143.
\\
Kapelko, V.V., and Linkevich, A.D. (1996). {\it Phys. Rev. E},
{\bf 54}, 2802.
\\
Ma, Y.-q., Zhang, Y., Ma, Y.-g., and Gong, C. (1993). {\it Phys.
Rev. E}, {\bf 47}, 3985.
\\
Nishimori, H., and Nonomura, Y. (1996). {\it J. Phys. Soc. Japan},
{\bf 65}, 3780.
\\
Nobili, R. (1985). Schr\"{o}dinger Wave Holography in Brain
Cortex. {\it Phys. Rev. A}, {\bf 32}, 3618.
\\
Ohya, M. (1989). {\it Reports Math. Phys.}, {\bf 27}, 19.
\\
Peru\v s, M. (1996). {\it Informatica}, {\bf 20}, 173.
\\
Peru\v s, M. (1997). In {\it Mind Versus Computer}, M. Gams, M.
Paprzycki and X. Wu, eds., IOS Press, Amsterdam, p. 156.
\\
Peru\v s, M. (1998). {\it Zeitschr. angewan. Math. \& Mech.}, {\bf
78}, S 1, 23.
\\
Peru\v s, M. (2002). {\it Int. J. Computing Anticip. Sys.} {\bf
13}, 376-391. \\
Peru\v s, M., and Dey, S.K. (2000), {\it Appl. Math. Lett.} {\bf
13} (8), 31.
\\
Peru\v s, M., and E\v cimovi\v c, P. (1998). {\it Int. J. Appl.
Sci. Computat.} {\bf 4}, 283.
\\
Pribram, K.H. (1991). {\it Brain and Perception}, Lawrence Erlbaum
A., Hillsdale.
\\
Pribram, K.H., ed. (1993). {\it Rethinking Neural Networks
(Quantum Fields and Biological Data)}, Lawrence Erlbaum A.,
Hillsdale.
\\
Psaltis, D., Brady, D., Gu, X.-G., and Lin, S. (1990). Holography
in ANNs. {\it Nature}, {\bf 343}, 325.
\\
Psaltis, D., and Mok, F. (1995). {\it Scien. Amer.}, {\bf 273},
No. 5 (November), 52.
\\
Scarani, V. (1998). {\it Amer. J. Phys.}, {\bf 66}, 956.
\\
Schempp, W. (1994). In {\it Wavelets and Their Applications}, J.S.
Byrnes {\it et al.}, eds., Kluwer, Amsterdam, 1994, p. 213.
\\
Steane, A. (1998). {\it Reports on Progress in Phys.}, {\bf 61},
117.
\\
Sutherland, J.G. (1990). {\it Int. J. Neural Sys.}, {\bf 1}, 256.
\\
Vapnik, V.N. (1998). {\it Statistical Learning Theory}; sec.
11.12.1: Reproducing Kernels Hilbert Spaces. John Wiley \& Sons,
New York. \\
Weinacht, T.C., Ahn, J., and Bucksbaum, P.H. (1999). {\it Nature
(Letters)}, {\bf 397}, 233.
\\
Zak, M., and Williams, C.P. (1998). Quantum Neural Nets. {\it Int.
J. Theor. Phys.}, {\bf 37}, 651.
\\
Zurek, W.H. (1996). {\it Physics Today}, {\bf 44}, no. 10 (Oct.),
36.
\end{document}